# High Pressure Structural Stability of Multiferroic Hexagonal *RE*$MnO_3$

P. Gao[1], Z. Chen[1], T. A. Tyson[1,2], T. Wu[1], K. H. Ahn[1,2], Z. Liu[3], R. Tappero[4]
S. B. Kim[2] and S.-W. Cheong[2]
[1]Department of Physics, New Jersey Institute of Technology, Newark, NJ 07102
[2]Rutgers Center for Emergent Materials and Department of Physics & Astronomy, Rutgers University, Piscataway, NJ 08854
[3]Geophysical Laboratory, Carnegie Institution of Washington, DC 20015
[4]National Synchrotron Light Source, Brookhaven National Laboratory, Upton, NY, USA

## Abstract

Structural changes in *RE*$MnO_3$ (*RE*= Y, Ho, Lu) under high pressure were examined by synchrotron x-ray diffraction methods at room temperature. Compression occurs more readily in the *ab* plane than along the *c*-axis. With increased pressure, a pressure-induced hexagonal to orthorhombic phase transition was observed starting at ~ 22 GPa for Lu(Y)$MnO_3$. When the pressure is increased to 35 GPa, a small volume fraction of Lu(Y)$MnO_3$ is converted to the orthorhombic phase and the orthorhombic phase is maintained on pressure release. High pressure IR absorption spectroscopy and Mn K-edge near edge x-ray absorption spectroscopy confirm that the hexagonal $P6_3cm$ structure is stable below ~20 GPa and the environment around Mn ion is not changed. Shifts in the unoccupied *p*-band density of states with pressure are observed in the Mn K-Edge spectra. A schematic pressure-temperature phase diagram is given for the small ion *RE*$MnO_3$ system.

# I. Introduction

The rare-earth (*RE*) manganites (such as $REMnO_3$, $REMn_2O_5$ and $RE_2Mn_2O_7$) are a group of systems with interesting properties which have attracted interest from the fundamental and basic physics perspectives [1]. The $REMnO_3$ systems fall into two basic structures: hexagonal and orthorhombic phase depending on the ionic radius of *r*. Small ionic radius $REMnO_3$ systems (*RE*= Ho-Lu, Y) are in the hexagonal structure and exhibits both ferroelectric and magnetic properties. The magnetic ordering temperature $T_N$ is ~90 K [2] and the ferroelectric ordering temperature $T_C$ is near ~900 K. The low-high temperature phase transition is predicted to pass through several aristotype phases [3]. The hexagonal structure ($P6_3cm$, Z = 6) is built up of $MnO_5$ trigonal bipyramids, which are illustrated in Fig. 1. In the basal (*ab*) plane, the pyramids are linked at the base corners to construct a triangular lattice. The *RE* ions are located between these $MnO_5$ layers and are linked with oxygen atoms. Each consecutive (*ab* plane) layer of $MnO_3$ (Fig. 1(b)) is rotated by 180° about the *c*-axis. They are geometrically frustrated antiferromagnetic materials [4]. The $Mn^{3+}$ spins and the associated magnetic exchange are confined to the basal (*ab*) plane, and $Mn^{3+}$ magnetic ground state produces a 120° angle between spins on local Mn triangles.

A first order structural transition occurs near 1200 K on cooling from the high temperature centro-symmetric $P6_3mmc$ (Z=2) phase to an asymmetric low temperature $P6_3cm$ phase. The onset of the ferroelectricity with the spontaneous polarization $\boldsymbol{P_s}$ along *c*-axis [5] is observed to occur near $T_C$ ~900 K. If the high temperature is accompanied with high pressure (~4 GPa and ~1000 °C) [6, 7, 8], the hexagonal (Hex) compound will be converted into a corresponding orthorhombic (Ortho) phase, which has a higher density and is strongly Jahn-Teller distorted [9]. By other techniques such as chemical solution method [10] and epitaxial thin film growth techniques [11,12], the orthorhombic compound also can be stabilized. On an appropriate perovskite substrate, the hexagonal $REMnO_3$ is forced into the orthorhombic form [11]. In other words, substrate strain can stabilize the orthorhombic phase for small size ions. Vice versa, the substrate like $ZrO_2$ ($Y_2O_3$) (YSZ) can epitaxially growth Hex-$DyMnO_3$ and Hex-$TbMnO_3$ films, which are normally in the orthorhombic structure [12]. When the normal Hex-$REMnO_3$ films are deposited on the YSZ substrate, the lattice parameter *c* of the film in the out-of-plane direction is tuned by the film thicknesses [12]. If the cell is elongated along *c*-axis, it would enhance the spontaneous polarization value. In a relaxed Hex-*RE*$MnO_3$ film, the $T_N$ value is lower than the value of the bulk material [13].

Under high pressure conditions, the dielectric and magnetic properties of Hex-*RE*$MnO_3$ have been studied [14, 15, 16, 17, 18] up to 6 GPa. In those works, the hexagonal structure is not changed [15,19]. The $HoMnO_3$ spin rotation transition temperature ($T_{SR}$) decreases linearly with applied pressure at a rate of $dT_{SR}/dp = -2.05$ K/GPa [14]. The Mn magnetic moment is suppressed. At 10 K, the ordered magnetic moment values are reducing with rates $dM/dP = -0.35$ μB $GPa^{-1}$ in $YMnO_3$ and -0.08 μB/Mn $GPa^{-1}$ in $LuMnO_3$ [16,17]. The decrease in $LuMnO_3$ is due to an enhancement of the geometrical frustration effects on the triangular lattice. At the same time, the magnetic symmetry of the triangular AF state (the irreducible representation Γ2) remains unchanged. However, spin reorientation of Mn magnetic moments and a change in the symmetry of the AF structure occur in $YMnO_3$. A spin-liquid state due to magnetic frustration on the triangular lattice forms by Mn ions at normal pressure and $T > T_N = 70$ K, and an ordered triangular AF state occurs with the symmetry of the irreducible representation Γ1 arises at $T < T_N$. The high-pressure effect leads to a spin reorientation of Mn magnetic moments and a change in the symmetry of the AF structure, which can be described by a combination of the irreducible representations Γ1 and Γ2 [16, 17,18]. High pressure induces a spin-liquid phase in $YMnO_3$, coexisting with the suppressed long-range AF order. The spin liquid phase exhibits a temperature dependence distinctively different from short-range spin correlations seen at ambient pressure. Its formation occurs through an in-plane Mn-O bond symmetrization and results in reduced magnetoelastic coupling at high pressures [19].

Structurally, high pressure Raman spectroscopy reveals the possibility of a hexagonal-orthorhombic phase transition happening at room temperature induced by pressure [20]. In a Hex-$Ho_{0.8}Dy_{0.2}MnO_3$, the phase transition pressure is interpreted to occur at ~10 GPa by observation of Raman phonon changes. No combined systematic study of the structure of Hex-*RE*$MnO_3$ under pressure has been conducted on this system to date.

In this work, we present the hydrostatic and quasi-hydrostatic pressure effects on *RE*$MnO_3$ (*R*= Y, Ho, Lu) to explore how the structural, vibrational and optical properties change with pressure. To effect this, synchrotron based x-ray diffraction (XRD), x-ray spectroscopy (XAS) and infrared (IR) methods are utilized at room temperature (298 K). We find that, at room temperature, a pressure-induced a hexagonal-orthorhombic phase transition requires a pressure above ~22 GPa for Lu(Y)$MnO_3$ and results in only small fractional conversion of Lu(Y)$MnO_3$ to the orthorhombic phase. The hexagonal *RE*$MnO_3$ is stable below this pressure. By checking the dipole moment of a unit cell, we find that the spontaneous polarization is a constant value under hydrostatic pressures at least up to ~11 GPa where we have conducted detailed structural refinement. The IR absorption spectrum and x-ray near edge absorption spectrum confirm that $P6_3cm$ is a stable structure under ~20 GPa and the environment around the Mn ion is not changed. The temperature for the transition to the orthorhombic phase is strongly suppressed by the application of pressure.

## II. Experimental and Computational Methods

Finely ground *RE*$MnO_3$ polycrystalline powders made by the solid-state reaction method were sealed into a diamond anvil cell (DAC) with three to four ruby chips for pressure calibration. Monochromatic X-ray ($\lambda$ = 0.4066 Å) powder diffraction under high pressures was completed at the X17C beam line at the National Synchrotron Light Source (NSLS), in Brookhaven National Laboratory. A charge-coupled device (Mar, 2048 × 2048 pixels with 80 μm resolution) was used to record the 2-D diffraction rings. The integration of the data to yield intensity *vs*. 2θ diagrams was conducted by the program Fit2D [21]. Methanol/Ethanol/Water (16:3:1) mixture was used as the pressure medium in x-ray diffraction experiments, whose glass transition pressure is ~10.5 GPa [22]. The pressure dependent synchrotron infrared (IR) absorption spectra were measured at the U2A beam line, at NSLS. A vacuum infrared microscope provided resolution of 4 $cm^{-1}$ in far IR (100 – 650 $cm^{-1}$) range. It was equipped with a 3.5-micron Mylar beam splitter, a 40 mm working distant reflecting objective and a Si bolometer detector. CsI was used to dilute the *RE*$MnO_3$ powder and acted as the pressure medium in IR absorption measurements. The pressures generated by DAC were calibrated by the Ruby Fluorescence method. The transmission mode x-ray absorption measurements were performed at beam line X27A at the Mn K-Edge in a DAC utilizing a pair of partially perforated diamond anvils. The pressure calibration was based on the compression curves [23, 24] from the Fe XAFS spectra of a thin Fe metal layer loaded in the cell with the sample.

To determine the nature of the fastest varying IR modes, the eigenvectors and eigenvalues of the dynamic matrix were calculated following the approach taken previously in $HoMnO_3$ [25]. The structure as function of pressure was simulated in the density functional calculations using the projector augment wave approach [26] for $YMnO_3$ and $LuMnO_3$ (with no open *RE* *f*-shells).

## III. Results and Discussion

Diffraction patterns of *RE*$MnO_3$ are shown in Fig. 2(a) for the pressure range up to ~17 GPa. At ambient pressure, three hexagonal phases show similar diffraction patterns. The lattice parameters (*a* and *c*) are *a* = 6.11505 Å, *c* =11.3781 Å ($YMnO_3$); *a* =6.13450 Å, *c* =11.41790 Å ($HoMnO_3$); and *a* =6.04845 Å, *c* =11.38300 Å ($LuMnO_3$) at ambient pressure, were extracted by the Rietveld method [27]. These values agree with the published results [7] ($YMnO_3$ *a* = 6.14666 Å, *c* =11.4411 Å, $HoMnO_3$ *a* =6.13820 Å, *c* =11.4118 Å and $LuMnO_3$ *a* =6.03011 Å, *c* =11.3648 Å). When the crystal cell is compressed, the diffraction peaks shift toward high $2\theta$ value and the peak intensities are varied as expected. It is a continuous process with no additional peaks appearing in this pressure range (up to ~17 GPa in the figure). Pressure dependence of the volume was fit by the Murnaghan equation of state: $V(P) = V_0(1 + B_0' \frac{P}{B_0})^{-1/B_0'}$, where $V_0$, $B_0$ and $B_0$` are the volume for a solid, the zero pressure bulk modulus and the bulk modulus pressure derivative. In the hexagonal phase, the $B_0$ values are 112±8 GPa ($YMnO_3$), 130±7 GPa ($HoMnO_3$) and 109±12 GPa ($LuMnO_3$). For the typical Ortho-$LaMnO_3$, the values of $B_0$ and $B_0$` are 108 ± 2 GPa, 8.5 ± 0.4 [28]. The highly anisotropic compressibility of P$6_3$*cm* structure is indicated by the $B_0$` values 12, 14 and 26 respectively, compared to the P*bnm* orthorhombic structure. Fig. 2(b) and (c) illustrate the pressure dependent cell volume change and lattice *a*, *c* variations. In the *ab* plane, the cell is much easier to be compressed than along the *c* direction. The pressure dependent variation of *a* is about twice the change for *c*. The volume and structural parameters of $YMnO_3$ and $LuMnO_3$ are calculated by DFT method. The DFT results show a coincidence with the experimental result on the pressure dependent volume change. While the $YMnO_3$ system is well matched by the model, the $LuMnO_3$ with a closed 4*f* shell is not well modeled.

By neutron diffraction, the structure under pressure has been determined up to 6 GPa [16, 17]. Here it is found that the oxygen position parameters do not significantly vary with pressure. Moreover, for the general hexagonal ($P6_3cm$) $ABO_3$ (A = *RE*, B= Mn, Fe and Ga) system, replacement of the A and/or B site by ions of different radii does not significantly modify the atomic positions [8, 29, 30, 31]. Mn is approximately in the centre of its oxygen environment [32]. In the DFT simulations, the same trend is observed up to 10 GPa. The x-ray scattering factor of oxygen is small comparing to *RE* and Mn, so uncertainties will be high on the O atomic positions. Hence in our fit process, the oxygen atoms are fixed and only the heavy metallic atomic positions (*RE* and Mn) are adjusted. This assumption is justified by the above observations. Below 11 GPa, the structural fit shows that the atomic positions of Mn and *RE* hold the original positions: *RE*1 at 2*a* (0, 0, ~1/4) and *RE*2 at 4*b* (1/3, 2/3, ~1/4), Mn at 6*c* (~1/3, 0, ~0), which is plotted in Fig. 3(a). By comparing the distances (below 11 GPa) between the Lu and Mn ions (Mn-Lu1 and Mn-Lu2 distances) in Fig. 3(b), we found that those values are conserved. This implies that the heavy atoms framework is unchanged (and the buckling of the layers is maintained at least up to 11 GPa); the oxygen atoms will be embedded into this framework as under normal pressure conditions. Beyond 11 GPa, an accurate Rietveld fit is not feasible due to distortion of the peak shapes by the glass transition of the pressure medium (non-hydrostatic conditions). However, the diffraction peak positions which determine the lattice parameters remain stable.

To further explore the pressure dependence at higher pressure, the pressure dependent IR optical density (OD) spectra are conducted and displayed in Fig. 4(a), 4(b), 4(c). The OD spectra of pure CsI have no significant change with the pressure increase (except the low frequency ~100 $cm^{-1}$ edge shift). Under the hydrostatic pressure condition, there is no abrupt profile shape change in the OD spectra. The phonons of the samples are hardening with pressures and the intensity is redistributed as expected. Natural $DyMnO_3$ is an orthorhombic phase and the OD spectrum is a typical example (Fig. 4 (d)). It is used as a reference to show that no conversion to this phase occurs to ~20 GPa. No characteristic peaks in common were found between the Hex-*RE*$MnO_3$ OD spectrum and the Ortho-$DyMnO_3$ OD spectrum. The spectrum shapes show a big difference in the region ~300-550 $cm^{-1}$ for the Ortho and Hex phases because of the structure difference of bipyramidal ($MnO_5$) and octahedral ($MnO_6$). Up to ~20 GPa, both the OD spectra and the X-ray diffraction result show no phase transition and the crystal structure is slightly modified. The $P6_3cm$ is a stable structure in this pressure range at room temperature. From that, we can deduce that in previous work of synthesis of meta-stable orthorhombic materials [6, 7, 8], the heat treatment plays the key role in the thermal dynamic process. With the thermal treatment (the temperature ~1000 K), the atoms acquire enough energy to overcome chemical barriers required to break bonds needed to construct a different symmetry under pressure (~4 GPa).

We now look in detail at the phonon pressure dependence. By the group theory analysis, the Hex-*RE*$MnO_3$ has been reported to have 23 IR active phonon modes (9A1+ 14E1) [33]. In our work, the ambient *RE*$MnO_3$ absorption spectrum shapes agree with the published results of $YMnO_3$ [33]. 11 Lorentzian oscillators [33] can index our ambient $YMnO_3$ OD spectrum well. The phonon assignments are listed in the Table 1 and compared to the measurements of Ref. [33]. The profiles are maintained for all IR pressures studied as pressure increases up to ~20 GPa. However, the asymmetry in compressibility in the *ab* plane *vs*. the *c* direction plays an important role in this variation between different phonons. Phonons shift to higher frequencies with differing rates resulting in pressure dependent changes in peak separation or even merger at specific pressures. For the $YMnO_3$ sample, at ~12 GPa, a weak phonon ~375 $cm^{-1}$ (labeled F) is separated from the phonon at ~351 $cm^{-1}$ which is the ambient pressure phonon ~310 $cm^{-1}$ (symmetry E1, labeled E), and the ambient pressure absorption lines of 265 $cm^{-1}$ (A1, labeled C) and 281 $cm^{-1}$ (E1, labeled D) merge into one at 301 $cm^{-1}$ at ~14 GPa. The OD spectra of Hex-$HoMnO_3$ and Hex-$LuMnO_3$ show more than 11 IR phonons at the ambient pressure. In Fig. 4(b) of the $HoMnO_3$ OD spectra, phonons ~468 $cm^{-1}$ (labeled I) and ~509 $cm^{-1}$ (labeled J) are indexed at ambient pressure. At ~10 GPa, the phonon ~457 $cm^{-1}$ (labeled H) merges into the phonon ~468 $cm^{-1}$ (label I); phonons ~ 492 $cm^{-1}$ and ~509 $cm^{-1}$ (near J) become one; When the pressure reaches ~16.5 GPa, two phonons with energy 256 $cm^{-1}$ (ambient, label B) and 268 $cm^{-1}$ (ambient, label C) merge. In Fig. 4(c) of the $LuMnO_3$ sample, weak vibrations with a maximum at ~416 $cm^{-1}$ (labeled H) and ~447 $cm^{-1}$ (near label I) and 515$cm^{-1}$ (~1.74 GPa, labeled K) can be found. Phonons at ~503 $cm^{-1}$(~~1.74 GPa), ~515 $cm^{-1}$ (label K) become one again at ~10.9 GPa; Near ~14.6 GPa, the ambient vibrations 268 $cm^{-1}$ (labeled B) and 286 $cm^{-1}$ (labeled C) are merging into ~293 $cm^{-1}$. The phonon with a frequency ~400 $cm^{-1}$ (ambient, labeled F) is two phonons ~432 $cm^{-1}$ and ~456 $cm^{-1}$ (labeled G) at ~20.5 GPa.

The pressure dependent phonon frequencies are plotted in Fig. 5. Systematically, the phonons harden with the pressure. The phonons in $LuMnO_3$ and $HoMnO_3$ are more sensitive to pressure than phonons in $YMnO_3$. For the three samples, the phonons with a large hardening rate occur at ~310 $cm^{-1}$ (labeled E) with 2.74 $cm^{-1}$/GPa ($YMnO_3$), 313$cm^{-1}$ (labeled E) with 3.2 $cm^{-1}$/GPa ($HoMnO_3$), and 332 $cm^{-1}$ (labeled E) with 2.9 $cm^{-1}$/GPa ($LuMnO_3$). They are assigned to E1 mode [33], which is corresponding to the motions of atoms O1, O2, O3 along a positive direction and atom O4, Mn along a negative direction in the *ab* plane. By calculating the E1 symmetry phonon modes for $LuMnO_3$ near ~320 $cm^{-1}$ by DFT method, we find that these strong pressure sensitive modes are mainly due to the motion of oxygen atoms. The oxygen atoms have significant oscillatory amplitudes with O ions moving in *ab* plane under high pressure. Fig. 6 displays the calculated displacement amplitudes and directions as vectors on O atoms for E1 (~320 $cm^{-1}$) symmetry phonon modes in ambient pressure $LuMnO_3$.

To verify above results, the pressure dependent near edge x-ray absorption spectrum measurement on $LuMnO_3$ were conducted. The Mn K-edge near edge x-ray absorption spectrum of Hex-$LuMnO_3$ was measured as a function of pressure up to 16 GPa in a diamond anvil pressure cell, which is shown in Fig. 7. No changes in the overall shape of profile were observed indicating no significant change in the space group up to this pressure. However, an expected shift in the main features is observed (the main "1*s* to 4*p*" peak move up by ~1.5 eV with between ambient and 16 GPa pressure). The shift in the peak position can be explained by the "Natoli Rule" [34] given by $(E_r\text{-}E_b)r^2$= constant, where $E_r$ is the resonance positions (4*p* or higher p-state here) and $E_b$ is a reference bound state position (pre-edge here). Compression of a coordination shell of radius *r* about Mn will shift the resonance position to higher energies. This is what is observed. The results are consistent with the Mn K-Edge measurements of Hex-$ScMnO_3$ [8] (a = 5.8308 Å and c = 11.1763 Å) compared to Hex-$YMnO_3$ [8] (6.1483 Å and c= 11.4131 Å) revealing a shift of the Sc system to features to higher energy compared to that for the Y system [35]. In both cases, the gross symmetry of the structure is preserved as can be seen in the overall profile shape [34].

To explore possible temperature dependence, we look at the OD spectra at fixed pressure but varying temperatures. The temperature was kept below 800 K [36] to avoid damage to the diamond cell resulting from loss of the adhesive holding the diamond anvils to the seats. The OD spectra in Fig. 8 were taken for temperatures up to 673 K with pressure kept at ~10 GPa. No phase change was observed to occur during the heating process. The spectra have the same peak positions and peak profiles after cooling the Hex-$LuMnO_3$ sample down to the room temperature. This again points to the fact that high temperature is the driving mechanism for overcoming the barrier.

The high temperature parent phase (P6/*mmc*) has a total dipole moment of zero. For the ferroelectric $P6_3cm$ structure, the dipole moment is not zero and is along the *c*–axis only. Given the atomic positions, the polarization can be calculated from the unit cell dipole moment ( $\vec{P} = \sum_i \vec{x}_i Q_i$ ). We used the formal charge values instead of born effective charges to estimate polarizations. In the Hex-*RE*$MnO_3$, the born effective charge is found to be close to the formal charge [37]. The polarizations in Table 2 were calculated for the ambient pressure structure, where the experimental polarization amplitude of Hex-$YMnO_3$ is 5.5 $\mu C/cm^2$ [38]. We plot the pressure dependent polarization of Hex-$LuMnO_3$ sample in Fig. 9, which are estimated from the fit structural parameters. Since the $P6_3cm$ structure persists under pressures and the atoms hold their fractional positions, the polarization shows approximately a constant in the hydrostatic pressure range up to ~11 GPa.

We now explore changes at higher quasi-hydrostatic pressures (up to 35 GPa) by x-ray diffraction method again. The Hex-*RE*$MnO_3$ (002) peak persists in diffraction patterns of $LuMnO_3$ and $YMnO_3$ samples at all pressures. At ~23 GPa, the pressure induced peaks appear in the x-ray diffraction patterns of $YMnO_3$ and $LuMnO_3$, which can not be indexed by the hexagonal structure. We attribute these peaks to the orthorhombic phase. The x-ray diffraction patterns are from a mixture of hexagonal and orthorhombic phases. At room temperature, only a small amount of the orthorhombic phase is converted and embedded into of Hex-*RE*$MnO_3$ grains for pressures up to ~35 GPa. Normalized high pressure diffraction patterns of $LuMnO_3$ and $YMnO_3$ are plotted in Fig. 10 (a) and (b). A simulation of the orthorhombic phase is given as the top curve in each figure. For $LuMnO_3$, the peaks appear at $2\theta = 6.8°$ and 10.7° at ~22 GPa. With pressure increasing to ~35 GPa, the peak intensities continuously grow. After releasing the pressure, these peaks remain. The pressure induced peaks are likely the peak (111) at $2\theta = 6.76°$ and the peak (022/211) at $2\theta = 11.3°$ of Ortho-$LuMnO_3$ at ambient condition. On the sample of $YMnO_3$, due to the different chemistry, only one pressure induced peak appears ($2\theta = 10.94°$ at ~24 GPa) and is observed to increase with pressure. It is attributed to the Ortho-$YMnO_3$ peak (022/211).

Including data points from our measurements and the literature (given below), a qualitative pressure-temperature (P-T) phase diagram of the *RE*$MnO_3$ system (based on the classic $YMnO_3$) is plotted in the Fig. 11. At ambient pressure, below ~1100 K, the system is in the $P6_3cm$ phase [3]. $YMnO_3$ will start to decompose when the temperature reaches ~1200 K [39,40]. Based on group theory arguments, the aristotype phases is predicted to be will be $P6_3/mcm$ at 1105 – 1360 K (intermediate phase), $P6_3/mmc$ at 1360 – 1600 K and possibly $P6/mmm$ above 1600 K [3]. To form Ortho-$YMnO_3$, the lowest pressure reported to date was ~3.4 GPa at ~920 K [41]. When temperature is higher than 1270 K [6,7,41,42], the required phase transition pressure is reduced significantly. Based on our fixed temperature and fixed pressure experiments, we deduce a mixed phase region of Hex-$YMnO_3$ and Ortho-$YMnO_3$ existing for temperatures below 920 K. Apart from the pressure and temperature, another important factor is the duration time for the Ortho-*RE*$MnO_3$ synthesis, which is not addressed here. Holding the Hex-$YMnO_3$ at high temperature and high pressure for a long time enhances the transformation.

## IV. Summary

In conclusion, we have explored the hexagonal *RE*$MnO_3$ structure under hydrostatic and quasi-hydrostatic pressures. The hexagonal form of *RE*$MnO_3$ is a very stable phase for pressures under ~20 GPa at ambient temperature as found from x-ray diffraction and x-ray absorption measurements. The unit cell is more readily compressed in the *ab* plane than along the *c*-axis. IR measurements reveal that the O atoms are the most sensitive to pressure and have significant oscillation amplitudes with O ions moving in *ab* plane. A phase transition to a perovskite structure is initiated above ~22 GPa. Above ~22 GPa, the hexagonal phase starts to be converted into the meta-stable orthorhombic (perovskite) phase, but the hexagonal phase is still the majority phase at ambient temperature.

## V. Acknowledgments

This research was funded by DOE Grant DE-FG02-07ER46402 (NJIT) and DE-FG02-07ER46382 (Rutgers). The U2A beam line at the National Synchrotron Light Source is supported by COMPRES, the Consortium for Materials Properties Research in Earth Sciences under NSF Cooperative Agreement EAR01-35554, U.S. Department of Energy (DOE-BES and NNSA/CDAC). Use of NSLS at Brookhaven

National Laboratory, was supported by the U.S. Department of Energy, Office of Science, Office of Basic Energy Sciences, under Contract No. DE-AC02-98CH10886. This research used resources of the National Energy Research Scientific Computing Center, which is supported by the Office of Science of the U.S. DOE under Contract No. DE-AC02-05CH11231.

**Table 1. IR Phonons of Hex-*RE*$MnO_3$ (*RE*=Y, Ho and Lu) at the ambient condition.**

| This work | | | YMnO$_3$ from Iliev [33] | | |
|---|---|---|---|---|---|
| YMnO$_3$ | HoMnO$_3$ | LuMnO$_3$ | | | |
| Expt. ($cm^{-1}$) | | | Expt. ($cm^{-1}$) | Mode symmetry | Direction and sign of the largest atomic displacements |
| 264.9 | 269.2 | 276.8 | 265 | *A1* | +z(*R*/Y1, *R*/Y2), -z(Mn) |
| 394.4 | 389.9 | 398.7 | 398 | *A1* | +z(O3), +z(O4), +x, y(O2), -x, y(O1) |
| 430.2 | 426.4 | 435.2 | 428 | *A1* | +z(O4,O3), -z(Mn) |
| | 467.6 | | | *A1* | +x, y(O1,O2), -x, y(Mn) |
| 619.1 | 614.5 | | 612 | *A1* | +z(O1,O2), -z(Mn) |
| 210.5 | 226.6 | 240.0 | 211 | *E1* | +x, y(O1,O2), -x, y(*R*/Y1,*R*/Y2) |
| 237.0 | 257.1 | 268.5 | 238 | *E1* | x,y(Mn,O3), z(O1,O2) |
| 280.2 | 296.7 | 304.7 | 281 | *E1* | +x, y(O1,O2), -x, y(O3) |
| 309.6 | 313.1 | 331.8 | 308 | *E1* | +x, y(O1,O2,O3), -x, y(O4,Mn) |
| | | 416.3 | | *E1* | +x, y(O1), -x, y(O2) |
| | | 447.7 | | *E1* | +x, y(O1), -x, y(O2) |
| 456.1 | 457.6 | 470.9 | 457 | *E1* | +x, y(O4,O3), -x, y(O2,O1,Mn) |
| 497.1 | 492.3 | 507.8 | 491 | *E1* | +x,y(O4,O3,O1,O2), -x, y(Mn) |
| | 508.9 | | | *E1* | x, y(O4) |
| 598.1 | 594.3 | 599.8 | 596 | *E1* | x, y(O3) |

**Table 2. The dipole moment and polarization of Y(Lu)$MnO_3$ in a $P6_3cm$ unit cell (ambient pressure)***

| | Dipole (*e*Å) | Polarization (μC/$cm^2$) |
|---|---|---|
| $YMnO_3$ | -2.16 | -5.78 [43] |
| | -2.06 | -5.51 [44] |
| | -1.75 | -4.69 [16] |
| $LuMnO_3$ | -2.71 | -7.55 [32] |
| | -3.89 | -10.81 [17] |

*Results calculated from crystal structure data in given references.

## Figure Captions

**Fig.1.** (Color online) (a) The crystal structure of hexagonal *RE*$MnO_3$ showing the $MnO_5$ bipyramids and (b) the triangular lattice of Mn ions linked by oxygen atoms.

**Fig.2.** (Color online) (a) X-ray powder diffraction patterns of *RE*$MnO_3$ at various pressures ($T$ = 298 K). From left panel to right panel are diffraction patterns of $YMnO_3$, $HoMnO_3$ and $LuMnO_3$, respectively. (b) Pressure dependent unit cell volumes (dots) and the first order Murnaghan fit (solid line) and (c) Compressibility of lattice parameters of *a, c and* V. The crosses are from the DFT calculations. Note the higher compressibility of the *a*-axis.

**Fig.3.** (Color online) Pressure dependent fractional positions of Lu1 2*a* (0, 0, *z*), Lu2 at 4*b* (1/3, 2/3, *z*) and Mn at 6*c* (*x*, 0, *z*) in (a) and the distance between Mn and Lu ions in (b).

**Fig.4**. (Color online) Pressure dependent IR optical density spectrum of Hex-*RE*$MnO_3$ at room temperature. (a) $YMnO_3$ IR absorption spectrum (Pressure: 0.00, 0.58, 1.17, 2.01, 3.02, 3.96, 4.98, 7.11, 7.97, 8.74, 9.35, 10.04, 10.99, 11.94, 12.98, 14.03, 15.08, 16.04, 17.11, 18.07, 19.40 GPa). (b) $HoMnO_3$ IR absorption spectrum (Pressure: 0.00, 0.58, 1.08, 2.01, 3.19, 4.13, 5.23, 5.91, 7.03, 8.14, 9.00, 10.29, 11.07, 12.56, 13.33, 14.03, 15.61, 16.40, 17.10, 18.25, 19.22, 20.46, 22.06 GPa). (c) $LuMnO_3$ IR absorption spectrum (Pressure: 0.00, 1.74, 2.95, 4.16, 5.30, 6.64, 7.99, 9.26, 10.88, 12.37, 14.04, 15.24, 17.31, 18.69, 20.65 GPa). (d) Ambient IR optical density spectra of Orthorhombic $DyMnO_3$.

**Fig.5.** (Color online) Pressure dependent phonon frequencies of $YMnO_3$, $HoMnO_3$ and $LuMnO_3$.

**Fig.6**. (Color online) Calculated E1 symmetry phonon modes for $LuMnO_3$ near ~320 $cm^{-1}$ showing the displacement amplitudes and directions as vector lengths and directions (can be positive or negative), respectively. For these modes with large pressure dependence only the oxygen atoms have significant oscillation amplitudes with O ions moving in *ab* plane.

**Fig.7.** (Color online) The Mn K-edge x-ray absorption spectrum of hexagonal $LuMnO_3$ at ambient pressure and 16 GPa. Note that the overall shape of the spectra is unchanged. Only a shift of feature positions due to compression is observed.

**Fig.8.** (Color online) Optical density spectrum of Hex-$LuMnO_3$ at RT, 373, 473, 573, 673K with the pressure at ~10 GPa, for each measurement.

**Fig.9.** (Color online) Calculated pressure dependent spontaneous polarization of Hex-$LuMnO_3$.

**Fig.10.** (Color online) X-ray diffraction patterns at quasi-hydrostatic pressures. Pressure induced peaks appear ~22 GPa ($LuMnO_3$) and ~24 GPa ($YMnO_3$). The arrow indicates induced peak positions and corresponding peaks in an orthorhombic phase; R means pressure release direction of measurement.

**Fig. 11.** (Color online) Qualitative Pressure-Temperature phase diagram for the *RE*$MnO_3$ system (based on the $YMnO_3$).

**Fig. 1. P. Gao *et al*.**

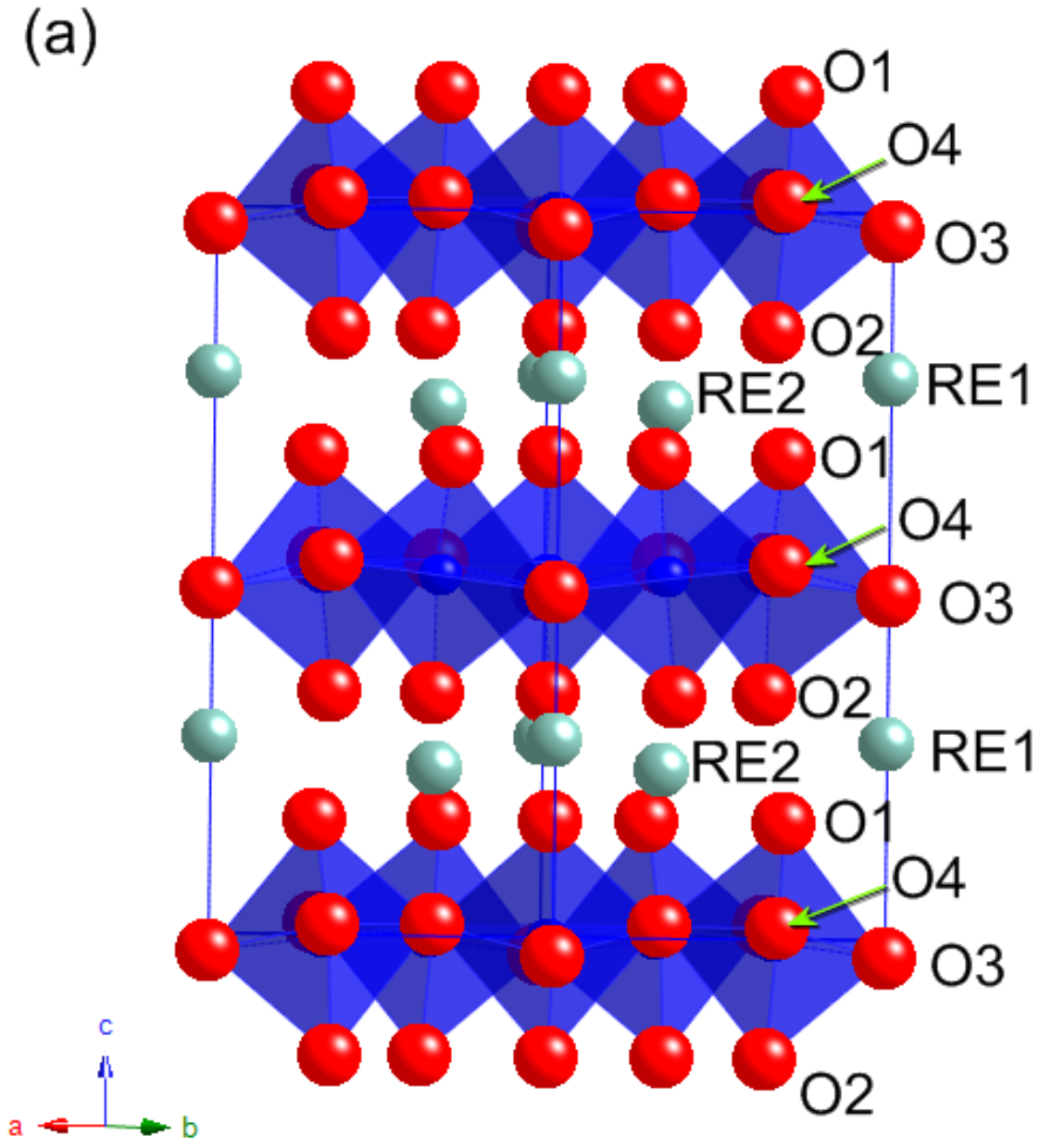

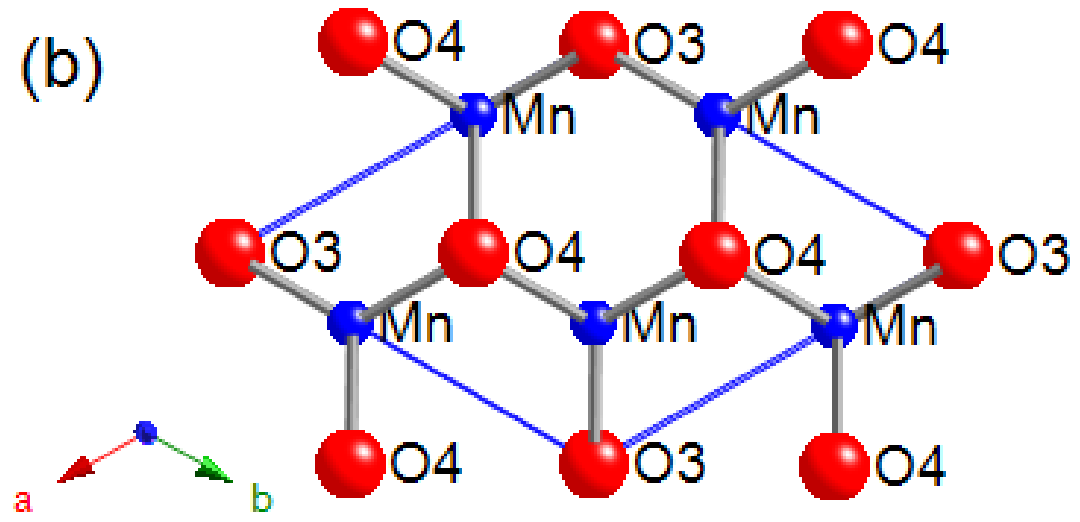

**Fig. 2. P. Gao *et al*.**

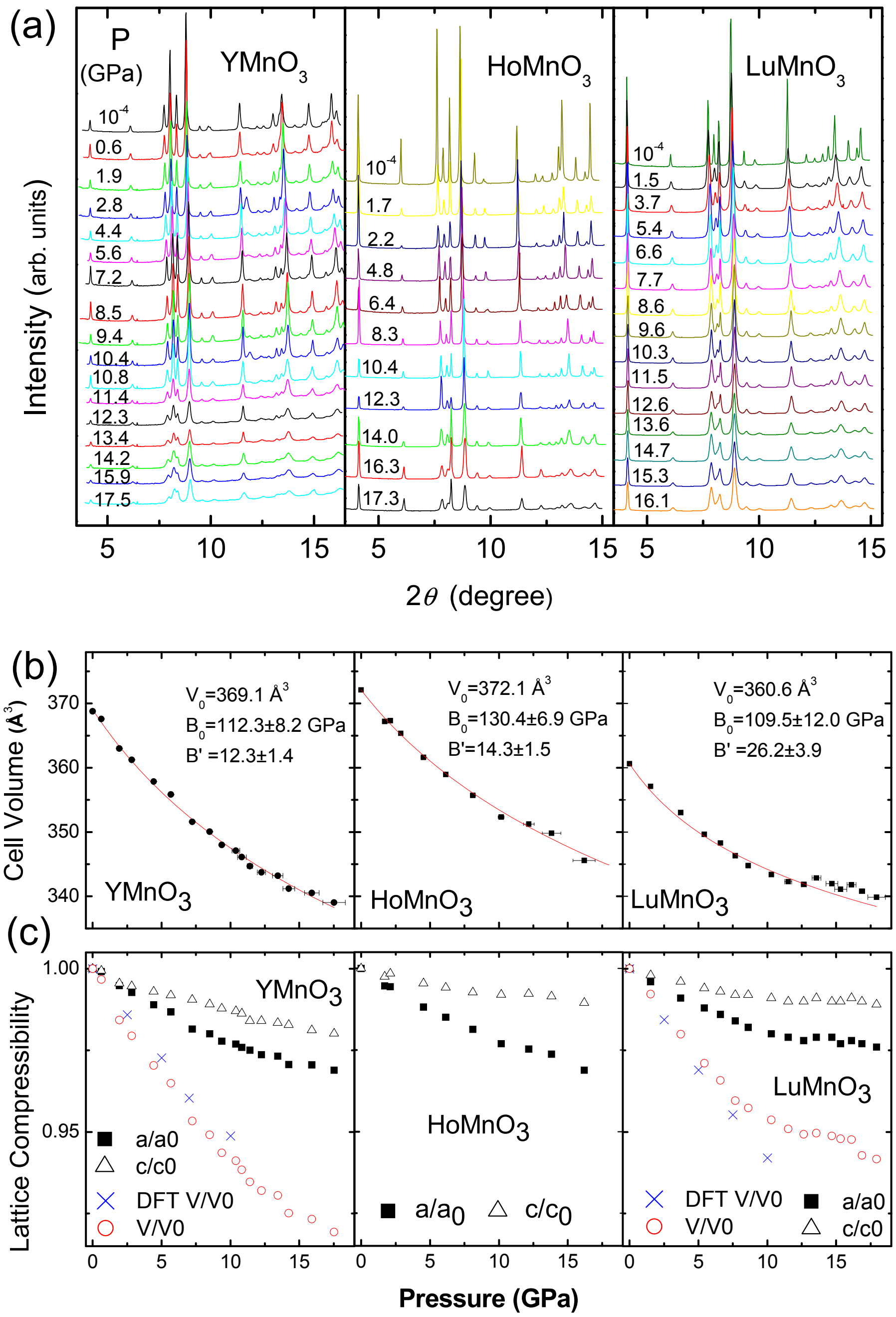

**Fig. 3. P. Gao *et al.***

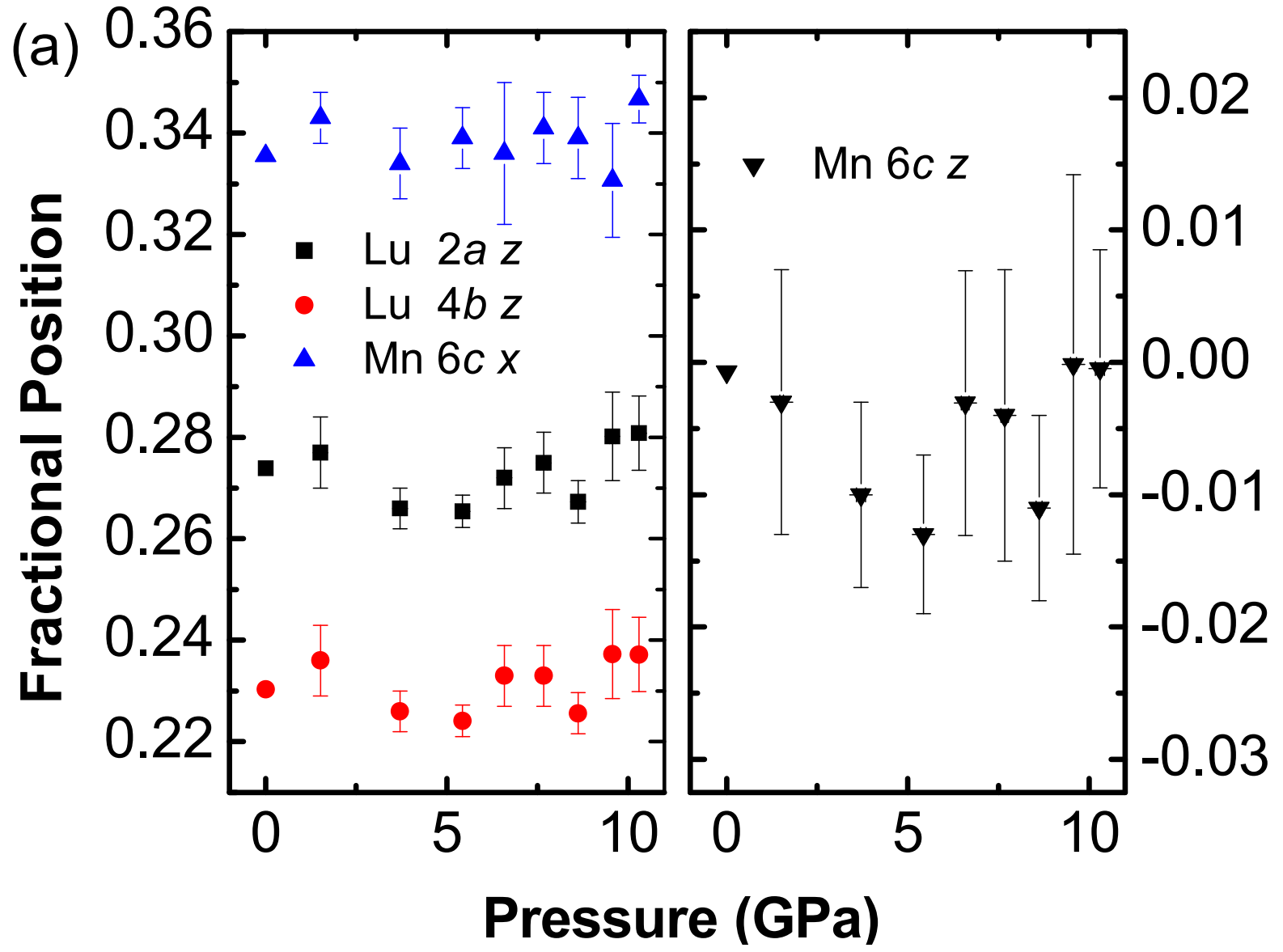

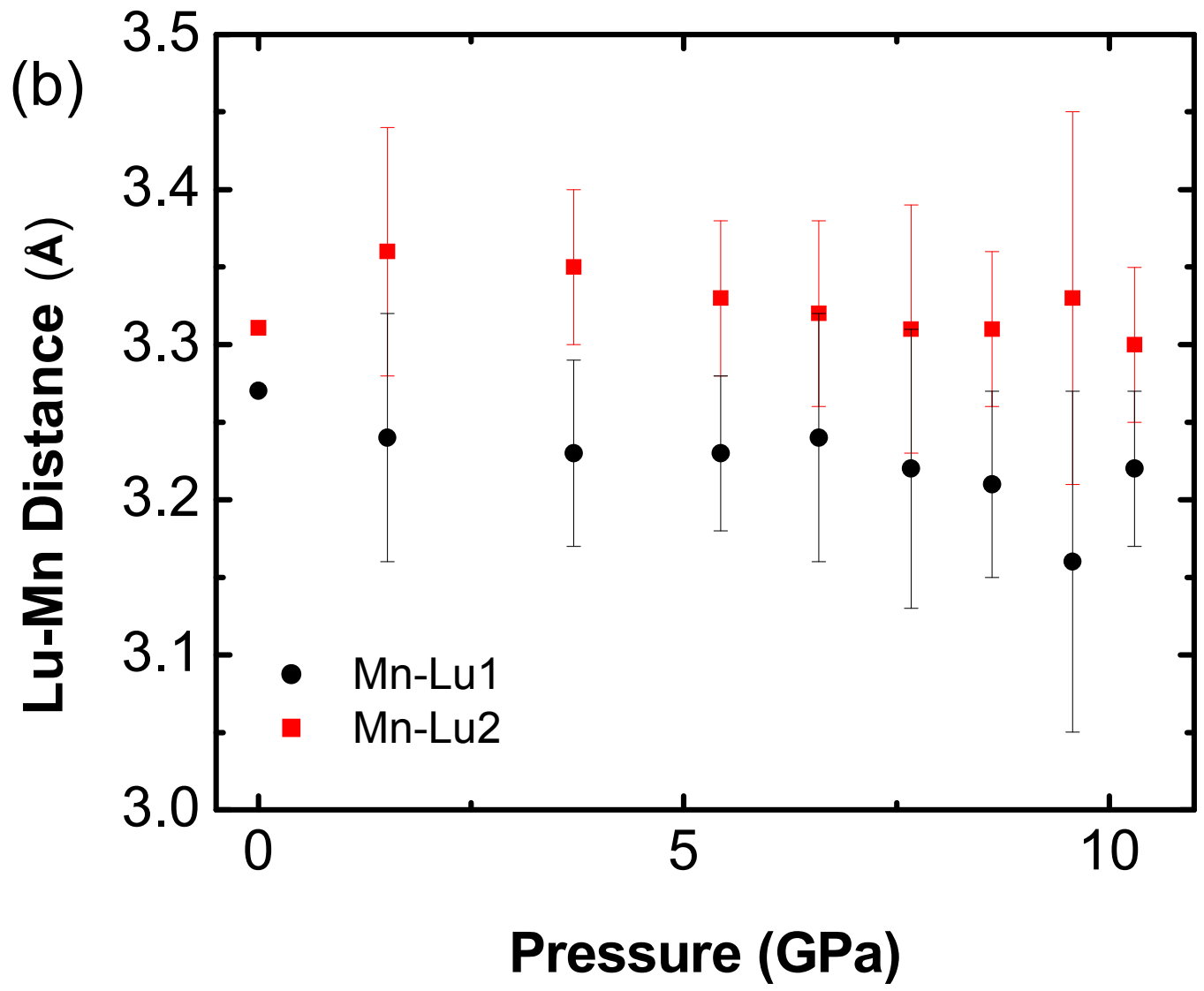

**Fig. 4. P. Gao *et al.***

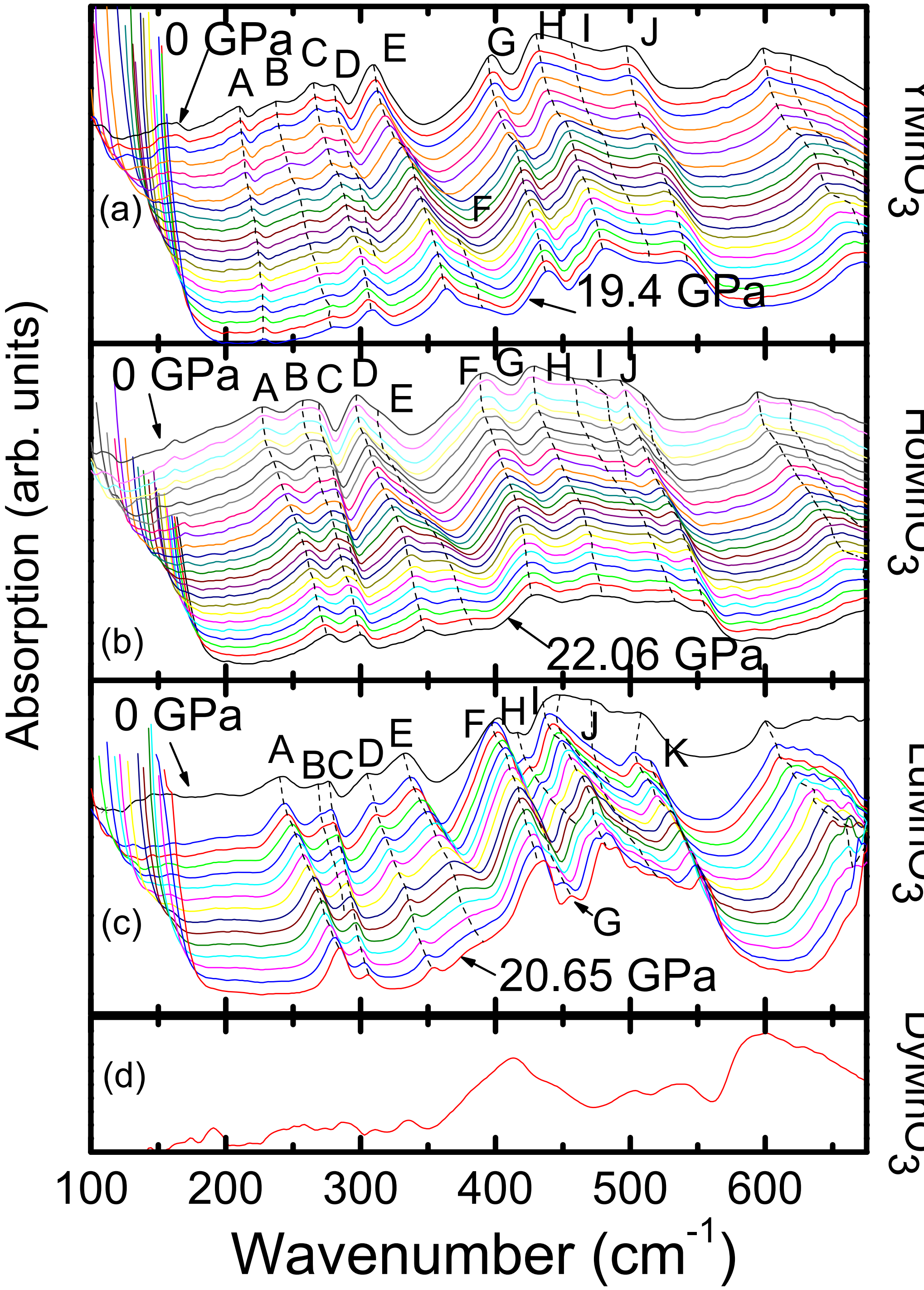

**Fig. 5. P. Gao *et al*.**

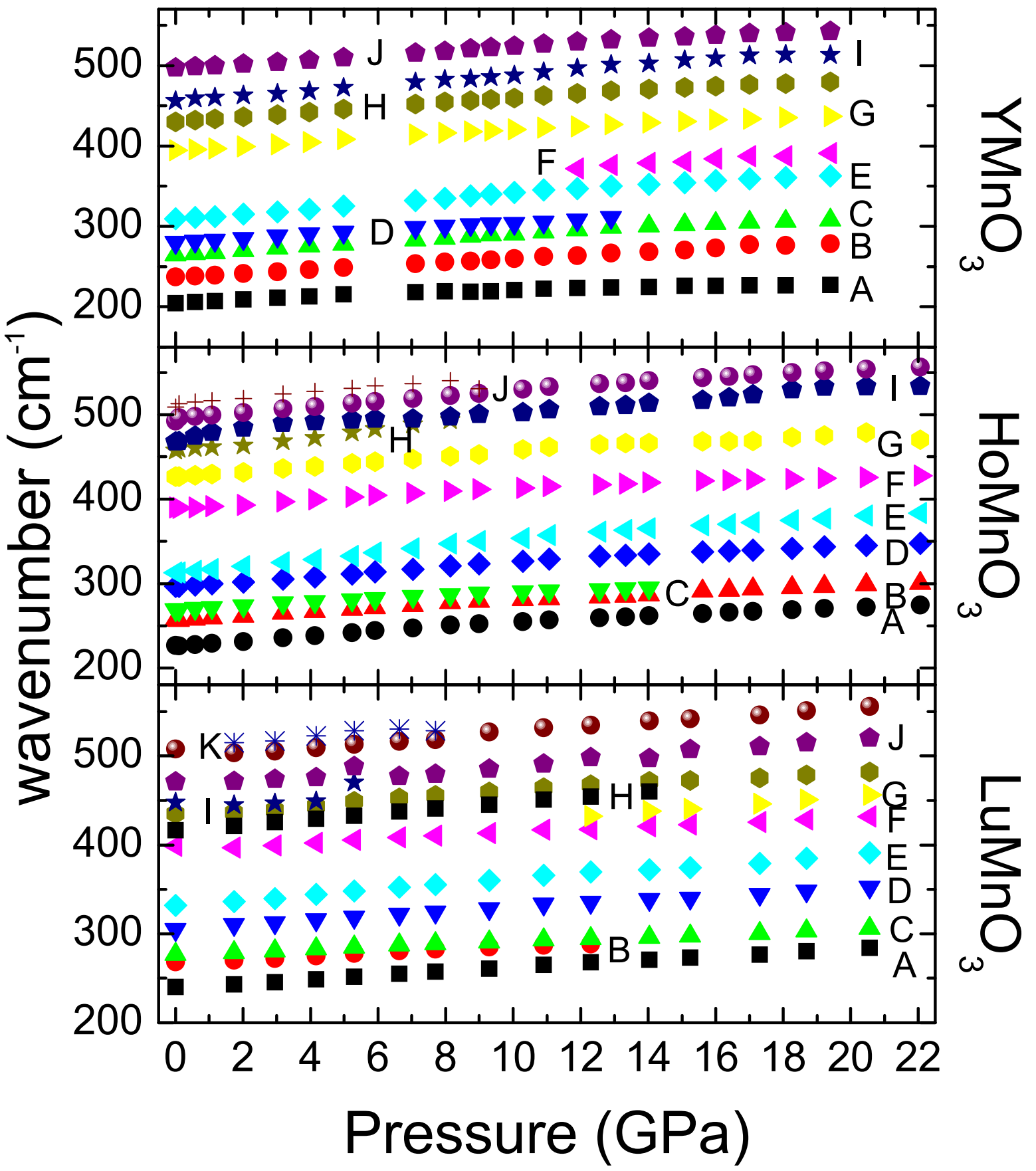

**Fig. 6. P. Gao *et al*.**

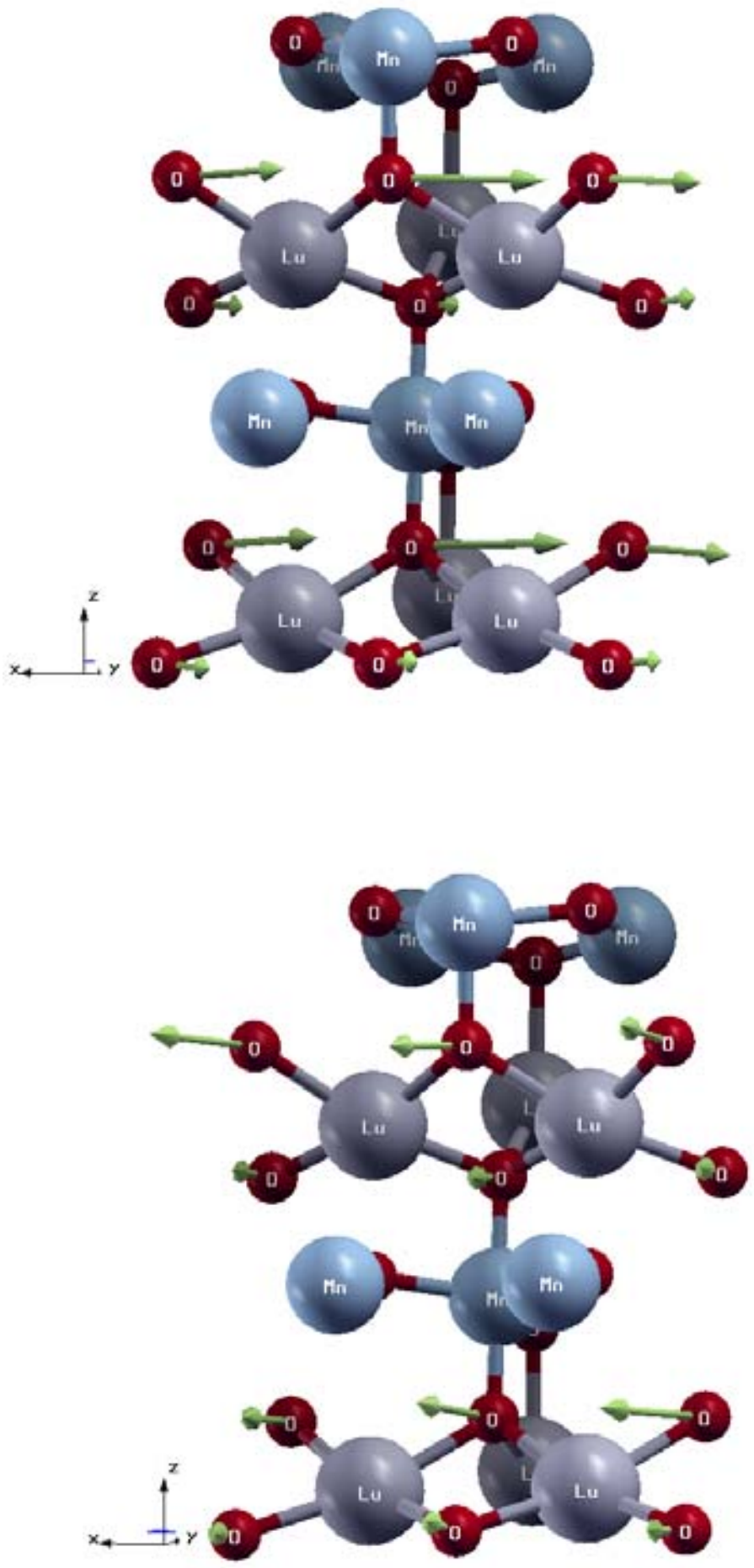

**Fig. 7. P. Gao *et al*.**

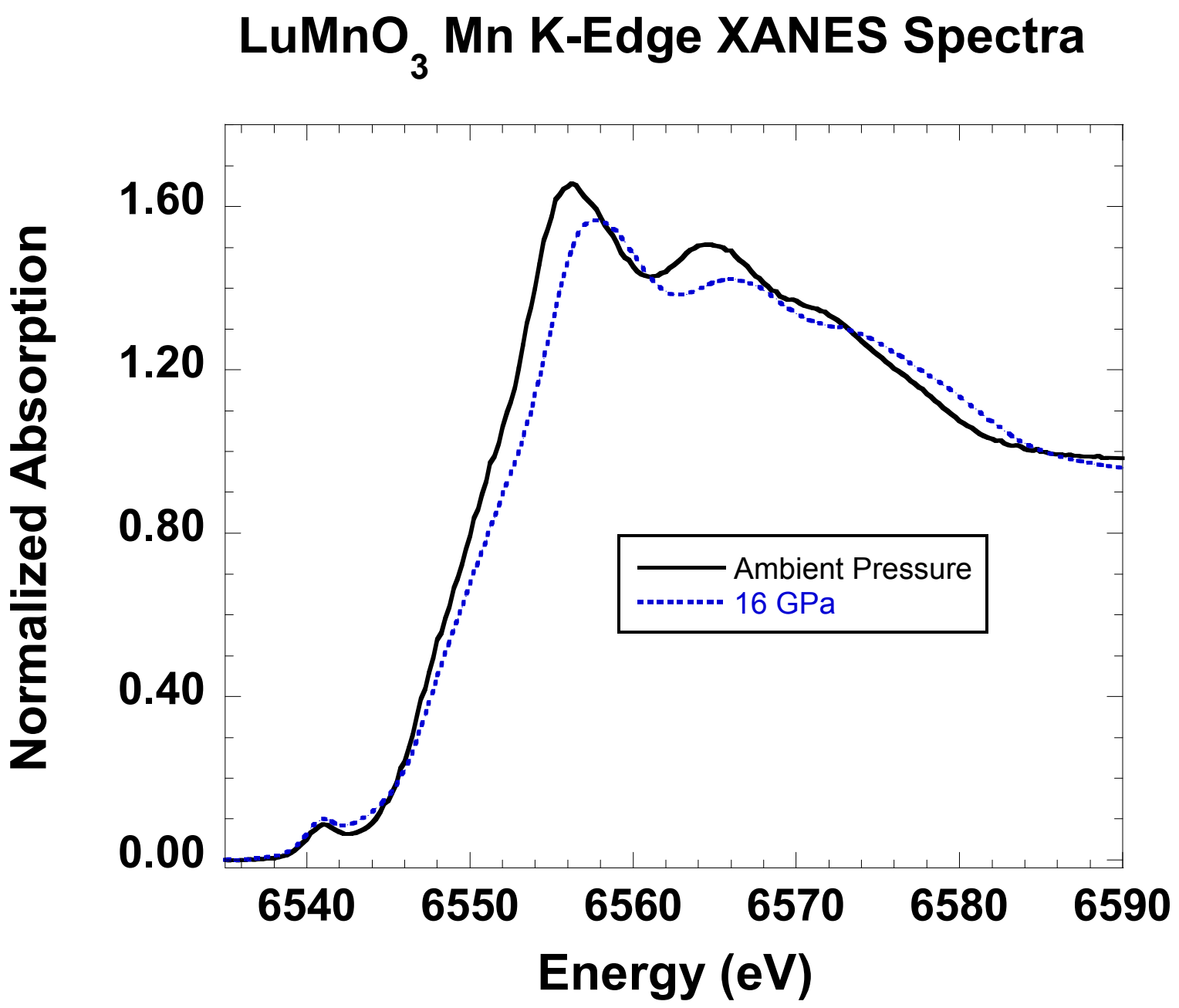

**Fig.8. P. Gao *et al.***

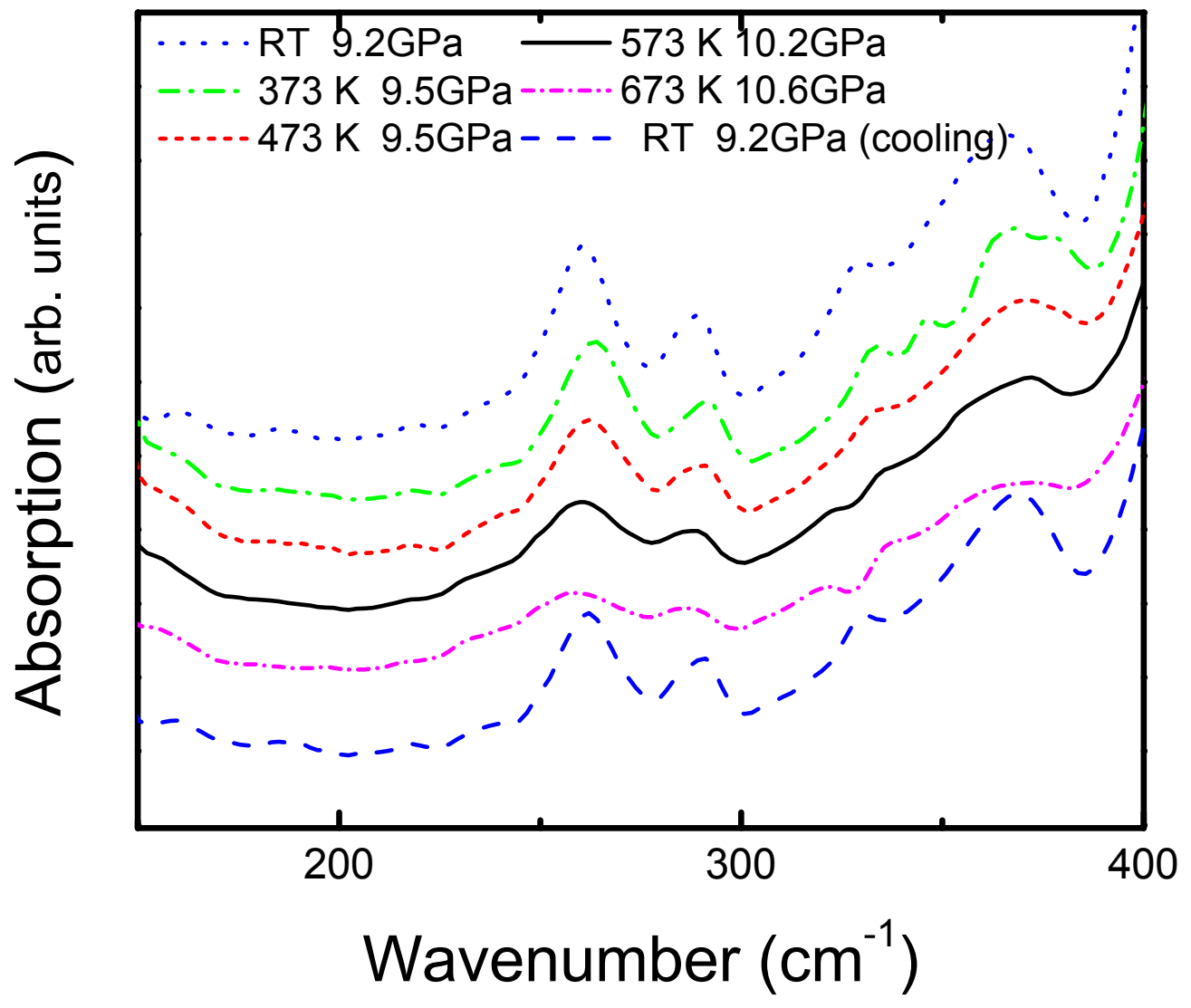

**Fig. 9. P. Gao *et al.***

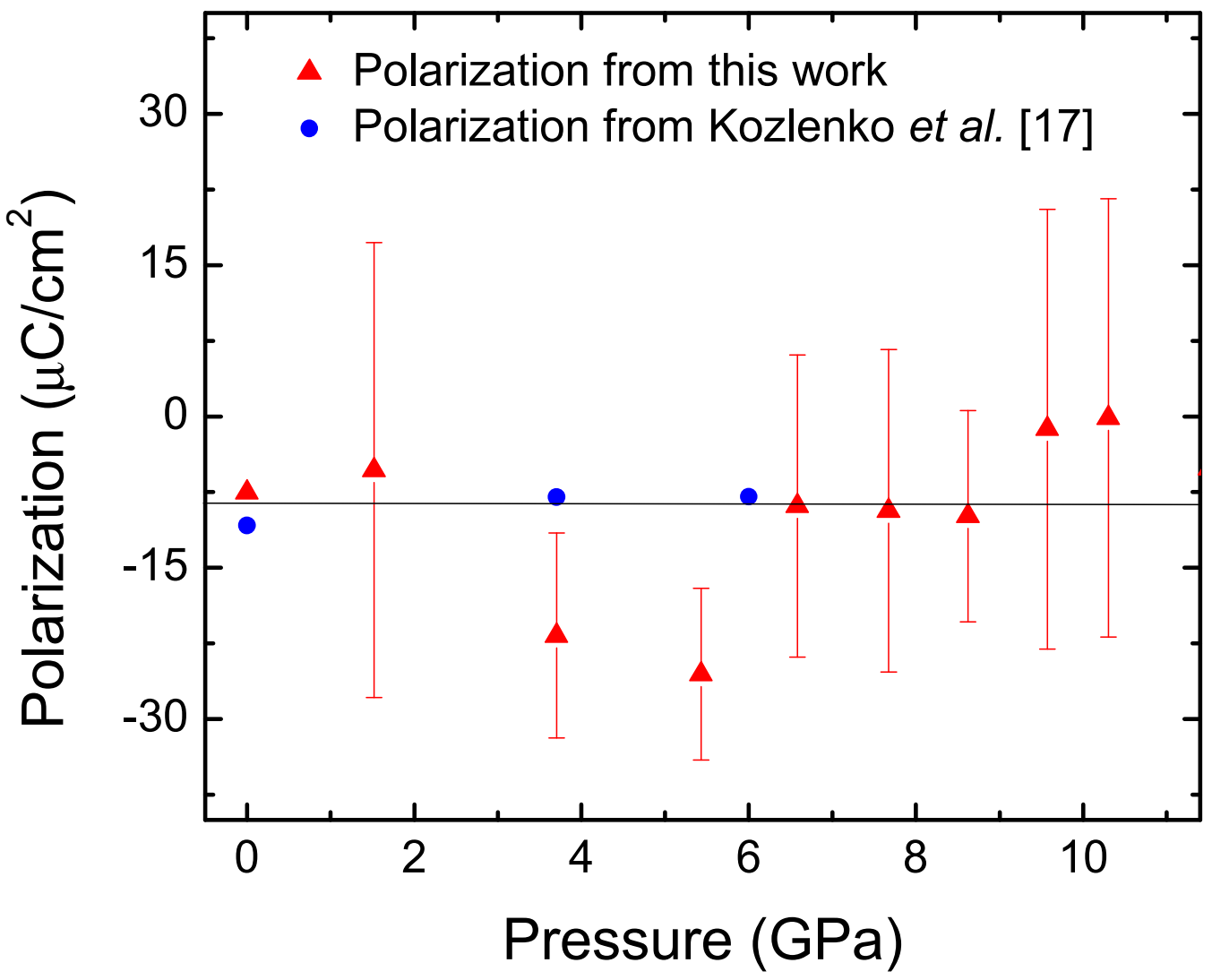

**Fig. 10. P. Gao *et al*.**

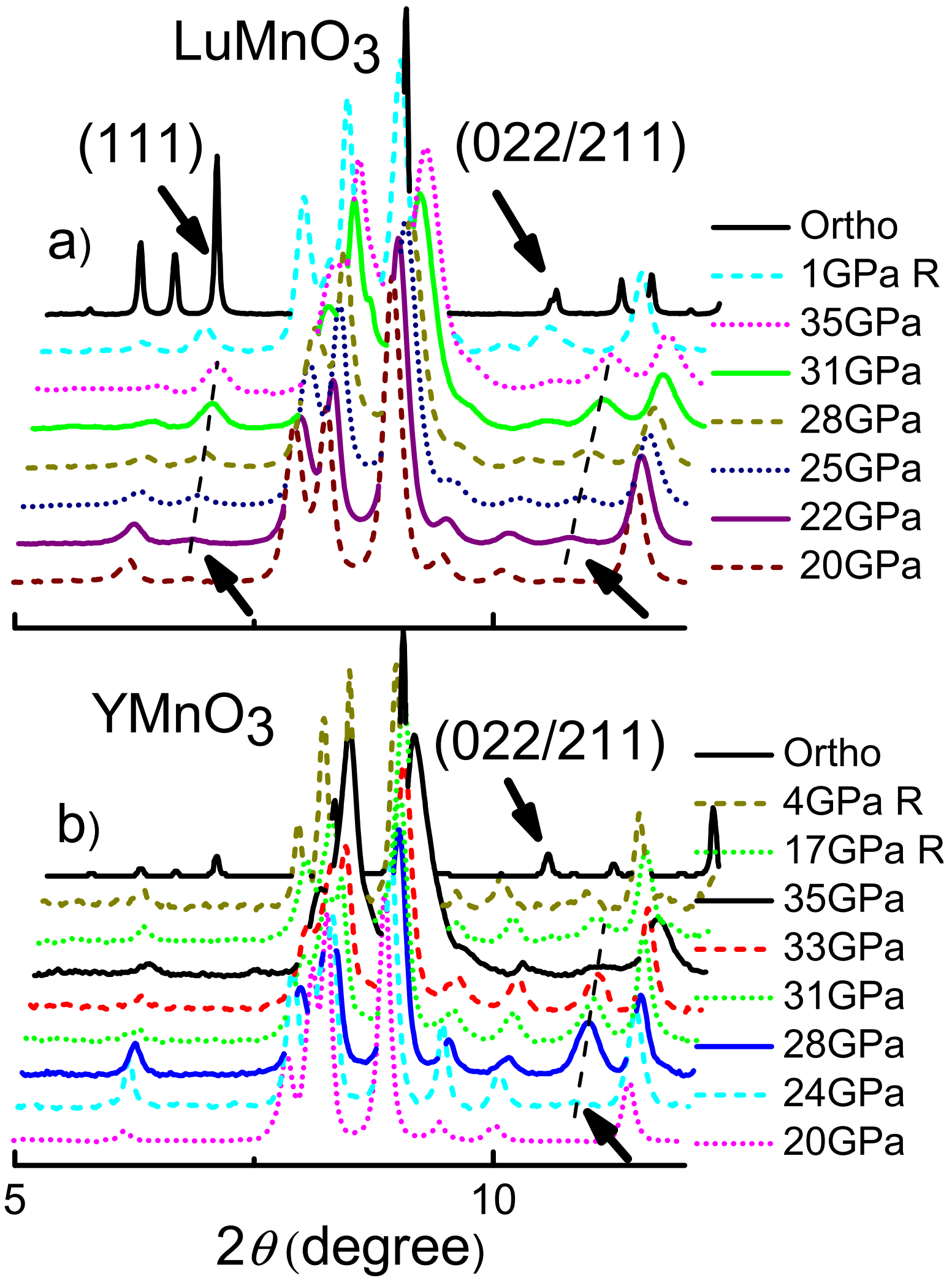

**Fig. 11. P. Gao *et al*.**

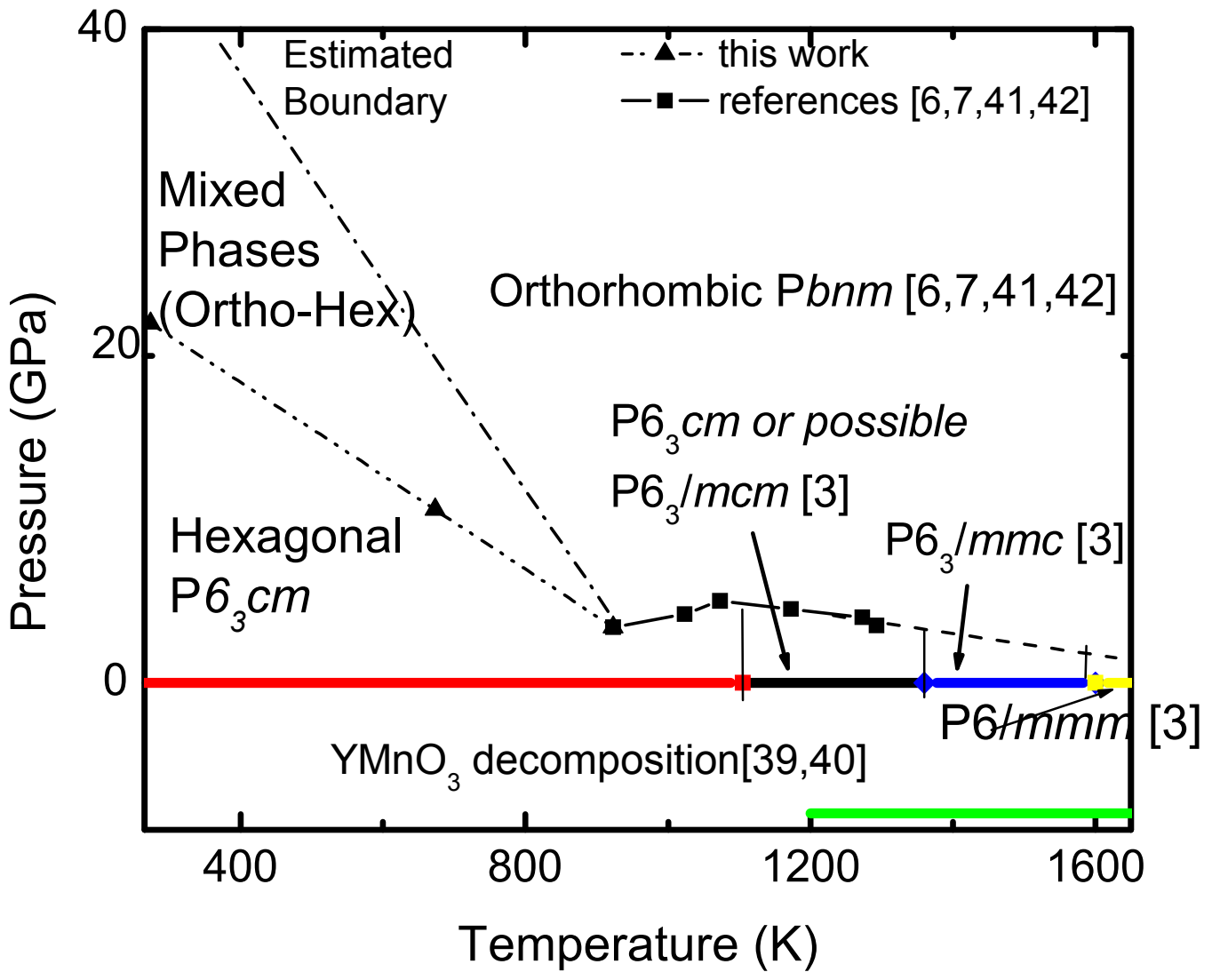

## References

1. K. F. Wang, J.-M. Liu, and Z. F. Ren, Adv. Phys. **58**, 321 (2009).
2. A. B. Souchkov, J. R. Simpson, M. Quijada, H. Ishibashi, N. Hur, J. S. Ahn, S. W. Cheong, A. J. Millis, and H. D. Drew, Phys. Rev. Lett. **91**, 027203 (2003).
3. S. Abrahams, Acta Crystallogr., Sect. B **65**, 450 (2009).
4. O. P. Vajk, M. Kenzelmann, J. W. Lynn, S. B. Kim, and S. W. Cheong, Phys. Rev. Lett. **94**, 087601 (2005).
5. D. Fröhlich, S. Leute, V. V. Pavlov, and R. V. Pisarev, Phys. Rev. Lett. **81**, 3239 (1998).
6. B. Lorenz, Y. Q. Wang, Y. Y. Sun, and C. W. Chu, Phys. Rev. B **70**, 212412 (2004).
7. J. S. Zhou, J. B. Goodenough, J. M. Gallardo-Amores, E. Moran, M. A. Alario-Franco, and R. Caudillo, Phys. Rev. B **74**, 014422 (2006).
8. K. Uusi-Esko, J. Malm, N. Imamura, H. Yamauchi, and M. Karppinen, Materials Chemistry and Physics **112**, 1029 (2008).
9. H. Okamoto, N. Imamura, B. C. Hauback, M. Karppinen, H. Yamauchi, and H. Fjellvag, Solid State Commun. **146**, 152 (2008).
10. H. W. Brinks, H. Fjellvåg, and A. Kjekshus, J. Solid State Chem. **129**, 334 (1997).
11. P. A. Salvador, T.-D. Doan, B. Mercey, and B. Raveau, Chem. Mater. **10**, 2592 (1998).
12. C. Dubourdieu, G. Huot, I. Gelard, H. Roussel, O. I. Lebedev, and G. Van Tendeloo, Philos. Mag. Lett. **87**, 203 (2007).
13. I. Gelard, C. Dubourdieu, S. Pailhes, S. Petit, and C. Simon, Appl. Phys. Lett. **92**, 232506 (2008).
14. C. dela Cruz, F. Yen, B. Lorenz, Y. Q. Wang, Y. Y. Sun, M. M. Gospodinov, and C. W. Chu, Phys. Rev. B **71**, 060407 (2005).
15. M. Janoschek, B. Roessli, L. Keller, S. N. Gvasaliya, K. Conder, and E. Pomjakushina, J. Phys.: Condens. Matter **17**, L425 (2005).
16. D. Kozlenko, S. Kichanov, S. Lee, J. Park, V. Glazkov, and B. Savenko, JETP Lett. **82**, 193 (2005).
17. D. Kozlenko, S. Kichanov, S. Lee, J. Park, V. Glazkov, and B. Savenko, JETP Lett. **83**, 346 (2006).
18. D. Kozlenko, S. Kichanov, S. Lee, J. Park, V. Glazkov, and B. Savenko, Crystallography Reports **52**, 407 (2007).
19. D. P. Kozlenko, I. Mirebeau, J. G. Park, I. N. Goncharenko, S. Lee, J. Park, and B. N. Savenko, Phys. Rev. B **78**, 054401 (2008).
20. S. M. Feng, L. J. Wang, J. L. Zhu, F. Y. Li, R. C. Yu, C. Q. Jin, X. H. Wang, and L. T. Li, J. Appl. Phys. **103**, 026102 (2008).
21. A. P. Hammersley, S. O. Svensson, and A. Thompson, Nuclear Instruments and Methods in Physics Research Section A: Accelerators, Spectrometers, Detectors and Associated Equipment **346**, 312 (1994).

22. R. J. Angel, M. Bujak, J. Zhao, G. D. Gatta, and S. D. Jacobsen, Journal of Applied Crystallography **40**, 26 (2007).
23. F. M. Wang and R. Ingalls, Phys. Rev. B **57**, 5647 (1998).
24. W. A. BASSETT and E. HUANG, Science **238**, 780 (1987).
25. T. A. Tyson, T. Wu, K. H. Ahn, S. B. Kim, and S. W. Cheong, Phys. Rev. B **81**, 054101.
26. G. Kresse and D. Joubert, Phys. Rev. B **59**, 1758 (1999).
27. B. Toby, Journal of Applied Crystallography **34**, 210 (2001).
28. I. Loa, P. Adler, A. Grzechnik, K. Syassen, U. Schwarz, M. Hanfland, G. K. Rozenberg, P. Gorodetsky, and M. P. Pasternak, Physical Review Letters **87**, 125501 (2001).
29. A. A. Bossak, I. E. Graboy, O. Y. Gorbenko, A. R. Kaul, M. S. Kartavtseva, V. L. Svetchnikov, and H. W. Zandbergen, Chem. Mater. **16**, 1751 (2004).
30. S. Geller, J. B. Jeffries, and P. J. Curlander, Acta Crystallogr., Sect. B **31**, 2770 (1975).
31. S. Lee, A. Pirogov, M. Kang, K.-H. Jang, M. Yonemura, T. Kamiyama, S. W. Cheong, F. Gozzo, N. Shin, H. Kimura, Y. Noda, and J. G. Park, Nature **451**, 805 (2008).
32. B. B. Van Aken, A. Meetsma, and T. T. M. Palstra, Acta Crystallogr., Sect. E **57**, i101 (2001).
33. M. N. Iliev, H. G. Lee, V. N. Popov, M. V. Abrashev, A. Hamed, R. L. Meng, and C. W. Chu, Phys. Rev. B **56**, 2488 (1997).
34. C. R. Natoli, edited by K. O. Hodgson, B. Hedman and J. E. Penner-Hahn (Springer-Verlag, New York, 1984), p. 38.
35. C. T. Wu, Y. Y. Hsu, B. N. Lin, and H. C. Ku, Physica B **329-333**, 709 (2003).
36. A. Jayaraman, Reviews of Modern Physics **55**, 65 (1983).
37. T. Tohei, H. Moriwake, H. Murata, A. Kuwabara, R. Hashimoto, T. Yamamoto, and I. Tanaka, Phys. Rev. B **79**, 144125 (2009).
38. N. Fujimura, T. Ishida, T. Yoshimura, and T. Ito, Appl. Phys. Lett. **69**, 1011 (1996).
39. I. K. Jeong, N. Hur, and T. Proffen, Journal of Applied Crystallography **40**, 730 (2007).
40. M. Chen, B. Hallstedt, and L. J. Gauckler, J. Alloys Compd. **393**, 114 (2005).
41. B. B. Van Aken, Ph. D. dissertation., University of Groningen, 2001.
42. V. E. Wood, A. E. Austin, E. W. Collings, and K. C. Brog, J. Phys. Chem. Solids **34**, 859 (1973).
43. G. nert, M. Pollet, S. Marinel, G. R. Blake, A. Meetsma, and T. T. M. Palstra, J. Phys.: Condens. Matter **19**, 466212 (2007).
44. B. B. Van Aken, A. Meelsma, and T. T. M. Palstra, "Acta Crystallogr., Sect. C: Cryst. Struct. Commun. " **57**, 230 (2001).